\documentclass[5p,twocolumn]{elsarticle}
\usepackage{graphicx,latexsym}
\usepackage{dcolumn}
\usepackage{amssymb,amsmath,bm}
\usepackage{subfigure}
\usepackage{braket}
\usepackage{siunitx}
\usepackage{array}
\usepackage{float}
\usepackage[nopar]{lipsum}
\usepackage{caption}
\usepackage{multirow}
\usepackage{bigstrut}

\biboptions{sort&compress}

\usepackage[super]{nth}

\newcommand{\angstrom}{\text{\normalfont\AA}}
\usepackage{hyperref}
\hypersetup{
    pdfnewwindow=true,       
    colorlinks=true,         
    linkcolor=blue,          
    citecolor=blue,          
    filecolor=magenta,       
    urlcolor=black           
}

\usepackage[normalem]{ulem}

\def\sec#1{Sec.\ \ref{#1}}

\def\fig#1{Fig.\ \ref{#1}}
\def\tab#1{Tab.\ \ref{#1}}

\journal{}

\begin{document}

\begin{frontmatter}


\title{Enhanced ultraviolet absorption in BN monolayers caused by tunable buckling}

\author[a1,a2]{Nzar Rauf Abdullah}
\ead{nzar.r.abdullah@gmail.com}
\address[a1]{Division of Computational Nanoscience, Physics Department, College of Science,
             \\ University of Sulaimani, Sulaimani 46001, Kurdistan Region, Iraq}
\address[a2]{Computer Engineering Department, College of Engineering,
	\\ Komar University of Science and Technology, Sulaimani 46001, Kurdistan Region, Iraq}

\author[a3]{Botan Jawdat Abdullah}
\address[a3]{Physics Department, College of Science, Salahaddin University-Erbil, Erbil 44001, Kurdistan Region, Iraq}

\author[a4]{Chi-Shung Tang}
\address[a4]{Department of Mechanical Engineering,
	National United University, 1, Lienda, Miaoli 36003, Taiwan}

\author[a5]{Vidar Gudmundsson}
\address[a5]{Science Institute, University of Iceland, Dunhaga 3, IS-107 Reykjavik, Iceland}


\begin{abstract}

The optical properties of a hexagonal Boron Nitride (BN) monolayer across the UV spectrum are studied by tuning its planar buckling. The strong $\sigma\text{-}\sigma$ bond through sp$^2$ hybridization of a flat BN monolayer can be changed to a stronger $\sigma\text{-}\pi$ bond through sp$^3$ hybridization by increasing the planar buckling. This gives rise to the $s$- and $p$-orbital contributions to form a density of states around the Fermi energy, and these states dislocate to a lower energy in the presence of an increased planar buckling. Consequently, the wide band gap of a flat BN monolayer is reduced to a smaller band gap in a buckled BN monolayer enhancing its optical activity in the Deep-UV region.
The optical properties such as the dielectric function, the reflectivity, the absorption, and the optical conductivity spectra are investigated. It is shown that the absorption rate can be enhanced by $(12\text{-}15)\%$ for intermediate values of planar buckling in the Deep-UV region, and  $(15\text{-}20)\%$ at higher values of planar buckling in the near-UV region. Furthermore, the optical conductivity is enhanced by increased planar buckling in both the visible and the Deep-UV regions depending on the direction of the polarization of the incoming light. Our results may be useful for optoelectronic BN monolayer devices in the UV range including UV spectroscopy, deep-UV communications, and UV photodetectors.

\end{abstract}

\begin{keyword}
Boron Nitride monolayers \sep Planar buckling \sep DFT \sep Electronic structure \sep  Optical properties
\end{keyword}

\end{frontmatter}

\section{Introduction}
Low-dimensional materials have attracted researcher's attention since the advances of technology come with a variety of novel material features. Because of the outstanding features of graphene \cite{Novoselov666, ABDULLAH2020126807} and other two-dimensional (2D) materials \cite{Mak2016,PhysRevLett.108.155501, Mannix2018,  ABDULLAH2020126350, Zhu2015}, this subject has recently experienced fast growth. Interlayer 2D materials are useful for a variety of functions and applications according to their electrical, optical, mechanical, magnetic, thermal, and catalytic capabilities \cite{ABDULLAH2021110095, C5CS00758E, ABDULLAH2021106073, GUPTA201544}.

The hexagonal Boron Nitride (h-BN) monolayer is one of the 2D material belonging to the III-V group has attracted a lot of experimental and theoretical attention, due to its extraordinary features and unique prospective uses. The BN monolayer is generated via a variety of processes. The edge fringe contrast and moiré patterns are relevant criteria for detecting the number of layers in the sheets and their stacking orientation. They were made using a chemical solution-derived technique \cite{han2008structure}, and a mechanical peeling of huge boron nitride monolayers was achieved using low-energy ball milling, which yielded high-quality BN monolayers with outstanding yield and efficiency \cite{li2011large}. Borazane is used as a precursor in atmospheric pressure chemical vapor deposition on copper to produce BN monolayer \cite{stehle2015synthesis}.

Strain, doping, electric fields, heterostructures, and planar buckling tuning are just several of the ways applied to modify the physical characteristics of BN monolayers due to their wide electrical band gap. It is envisaged that any changes in the electrical structure of systems with doping in a 2D h-BN will tune the band gap. Depending on the size and the symmetry of a C doping of a BN monolayer the electronic gaps may match nearly the entire visible electromagnetic spectrum. Increased doping causes gaps in the IR range, whereas low concentration causes gaps in the UV region \cite{park2012magnetic}. Doping of a BN monolayer with Ge results in a noticeable decrease in the energy gap as well as a significant reduction in the optical conductivity, with only minor absorption for wavelengths ranging from the infrared to visible light \cite{oliveira2019electronic}. Doped of a BN monolayer with Sn transforms a wide band gap BN monolayer into a semi-metallic system and the absorption coefficient in the visible and infrared regions is found to have a finite value, despite the fact that pure BN monolayer has no absorption in those ranges \cite{khan2020electronic}. The band gap of a monolayer BN is reduced in the presence of a Si dopant due to structural changes such as an increased band length, band angle, and a lattice constant \cite{ajaybu2021first}.

A pure h-BN monolayer has an indirect band gap. A P and an As substitution for N, results in a band gap reduction of 19.50 percent and 22.29 percent, respectively. An electric field can considerably lower this value, indicating that a combination doping with an external electric field might be a highly effective technique to fine-tune the electrical structure of BN monolayers \cite{hoat2020reducing}.
The band gap of a h-BN monolayer may be altered by biaxially applied stresses, with the band gap decreasing as the tensile strain rises \cite{fujimoto2014electronic}. The band gap of BN nanosheets can be controlled by applying stress and strain on them. Stress and strain have minimal influence on the reflectivity spectrum, while shifting the energy gap direction according to optical simulations \cite{jalilian2016tuning}, and a BN monolayer can function as a direct band gap semiconductor under tensile strain \cite{ghasemzadeh2018strain}.

Graphene quantum dots have been inserted into a BN monolayer to modify their electrical and optical properties, and when the size of graphene dots is increased, the exciton binding energies drop \cite{ding2018tuning}. Graphene and BN monolayer stacked in different ways could open the band gap of graphene and the semiconducting heterostructures of periodic graphene/h-BN absorb light throughout a wide frequency range, from the near the IR to the UV, according to calculations of their optical spectra \cite{aggoune2020structural}. On the other hand, another interesting technique to tune and improve the electron and optical characteristics of a 2D material is tuning the planar buckling  \cite{jalilian2016buckling}. The optical spectrum is red-shifted, when $\sigma$ bonds weaken and the number of $\pi$ electrons increases, and the planar buckling can result in a tunable band gap as the energy band gap decreases, when the planar buckling parameter is increased, according to the findings.

Furthermore, Deep-ultraviolet (DUV) communications have recently attracted a lot of attention due to a number of advantages over visible light communications, as well as advances in light-emitting diodes, detectors, spectroscopy, and other optoelectronic devices  \cite{he20191, hartmann2016preparation, cheng2021enhanced}. A BN monolayer can be synthesized by high-temperature molecular beam epitaxy on graphite, with the material having high optical properties in the deep ultraviolet \cite{vuong2017deep, elias2019direct}. The structural and the optical characteristics of a hexagonal BN have been explored using density functional theory. Materials having a broad band gap absorb effectively in the UV region, with a maximum reflectance of about 10 percent in the DUV area between the energy ranges of 0 to 25 eV, but no absorption is seen in the visible range, according to an investigation \cite{satawara2021structural}.

In the present work, the electrical and the optical properties of a h-BN monolayer are studied for both flat and buckled structures using first-principles calculations within the context of density functional theory. The influence of planar buckling on the electrical and optical characteristics of a BN monolayer is investigated using a variety of planar buckling values.  The findings suggest that planar buckling can offer a tunable band gap, and that when the critical value is reached, planar buckling flips the energy band gap type from indirect to direct. On the other hand, when the planar buckling factor is raised all optical spectra are red-shifted to a lower energy, and the visible region of the electromagnetic absorption spectrum undergoes a change. Our research offers a new path for novel optical device functions in nanotechnology.

The computational methodologies and model structure are briefly discussed in \sec{Sec:Methodology}. The major achieved outcomes are examined in \sec{Sec:Results}. The conclusion of the results is reported in \sec{Sec:Conclusion}

\section{Methodology}\label{Sec:Methodology}
A two-dimensional metallic nitride monolayer, h-BN monolayer, consists of of $2\times 2$ supercell with equal number of B and N atoms. In the calculations we select cutoffs for the calculation of the charge densities and the kinetic energy of the plane-waves at $1.088 \times 10^{4}$~eV \cite{ABDULLAH2021413273}, and $1088.5$~eV, respectively. The BN monolayer is considered fully relaxed when all the forces on the atoms are less than $10^{-5}$ eV/$\angstrom$ in the $18 \times 18 \times 1$ Monkhorst-Pack grid.
The interaction of layers in the BN structure can be diminished by considering a large value for a  vacuum layer of $20 \, \angstrom$ in the $z$-direction \cite{ABDULLAH2020100740, ABDULLAH2020103282}.

The exchange and correlation terms are approximated by a generalized gradient approximation with the Perdew-Burke-Ernzerhof (GGA-PBE), which is implemented in Quantum espresso (QE) software \cite{Giannozzi_2009, giannozzi2017advanced}. The QE is based on Kohn-Sham density functional theory (KS-DFT). For both the Self-Consistent Field (SCF), and the non-self-consistent field (NSCF) calculations for the band structure and the density of states (DOS) of a BN monolayer, a Monkhorst-Pack grid of sizes $18 \times 18 \times 1$ and $100 \times 100 \times 1$ are used, respectively \cite{ABDULLAH2021106981}.
The QE is used to calculate the optical properties of the BN monolayers with an optical broadening of $0.1$~eV \cite{ABDULLAH2021114644}.

\section{Results}\label{Sec:Results}
In this section, we present the electronic and optical properties of a BN monolayer with different planar buckling strength, $\Delta$.

\subsection{Electronic properties}
The physical properties of both flat and buckled BN monolayers are investigated
using the results of a flat BN monolayer as reference points to compare to.
In a flat BN monolayer, all the B and the N atoms are situated in the same $xy$-plane.
Once there is a planar buckling, the B atoms are situated in the same plane
and all the N atoms are arranged in another plane. The planar bucking parameter is defined as the vertical distance, $\Delta$,
between the B and the N planes. We will consider seven values of $\Delta$ starting from $0.1$ to $0.7$~$\angstrom$ in addition to the flat BN monolayer with $\Delta = 0.0$,
and we will find later critical values and limitation to the planar buckling.
\begin{table}[h]
	\centering
	\begin{center}
		\caption{\label{table_one} The bond length of B-N, lattice parameters, $a$, band gap (E$_g$), and formation energy (E$_{f}$) for different values of planar buckling, $\Delta$.}
		\begin{tabular}{|l|l|l|l|l|l|l|l|}\hline
			$\Delta$	&  B-N ($\angstrom$) &  a ($\angstrom$)   &  E$_g$ (eV)   &   E$_f$ (eV) \\ \hline
			0.0	    & 1.447 & 2.507 &  4.63    &  -72.205      \\
			0.1	    & 1.451 & 2.507 &  4.569   &  -71.917      \\
			0.2	    & 1.461 & 2.507 &  4.261   &  -71.049      \\
			0.3	    & 1.478 & 2.507 &  3.799   &  -69.607      \\
			0.4	    & 1.501 & 2.507 &  3.30    &  -67.580      \\
			0.5	    & 1.531 & 2.507 &  2.30    &  -64.971      \\
			0.6	    & 1.567 & 2.507 &  1.106   &  -61.765      \\
			0.7	    & 1.697 & 2.507 &  0.0001  &  -58.094      \\ \hline
	\end{tabular}	\end{center}
\end{table}

\begin{figure}[htb]
	\centering
	\includegraphics[width=0.5\textwidth]{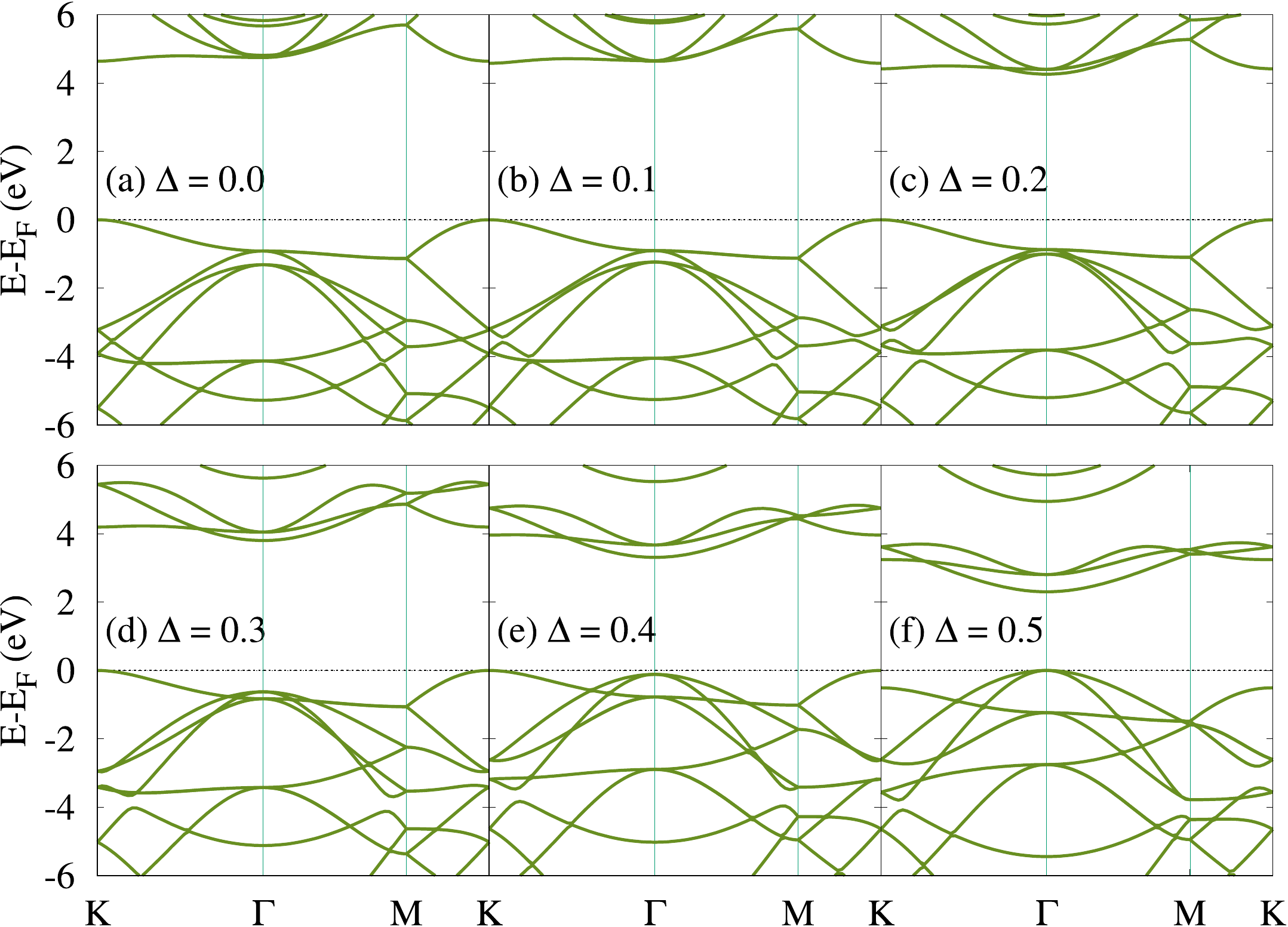}
	\caption{Band structure for optimized BN monolayers with planar buckling, $\Delta = 0.0$ (a), $0.1$ (b), $0.2$ (c), $0.3$ (d), $0.4$ (e), and $0.5$~$\angstrom$ (f).
		The energies are with respect to the Fermi level set to zero.}
	\label{fig01}
\end{figure}

We first study the energetic stability via the formation energy, E$_f$, which indicates
the structural stability of the buckled BN monolayers in comparison to a flat BN monolayer.
Our DFT calculations of the formation energies of the BN monolayers with different buckling are
shown in \tab{table_one}.
A BN monolayer with the selected values of $\Delta$ is quite stable up to $\Delta = 0.5$~$\angstrom$, but
our results demonstrate that the formation energy increases with increasing $\Delta$.
As the planar buckling increases, a BN monolayer is less energetically stable.
A BN monolayer with a low planar buckling, $\Delta = 0.1 \, \angstrom$, is the most stable structure among the buckled monolayers. Similar behavior was found for BeO monolayers. The stability of a BeO monolayer is decreases with increasing planar buckling \cite{jalilian2016buckling}.
\lipsum[0]
\begin{figure*}[htb]
	\centering
	\includegraphics[width=0.9\textwidth]{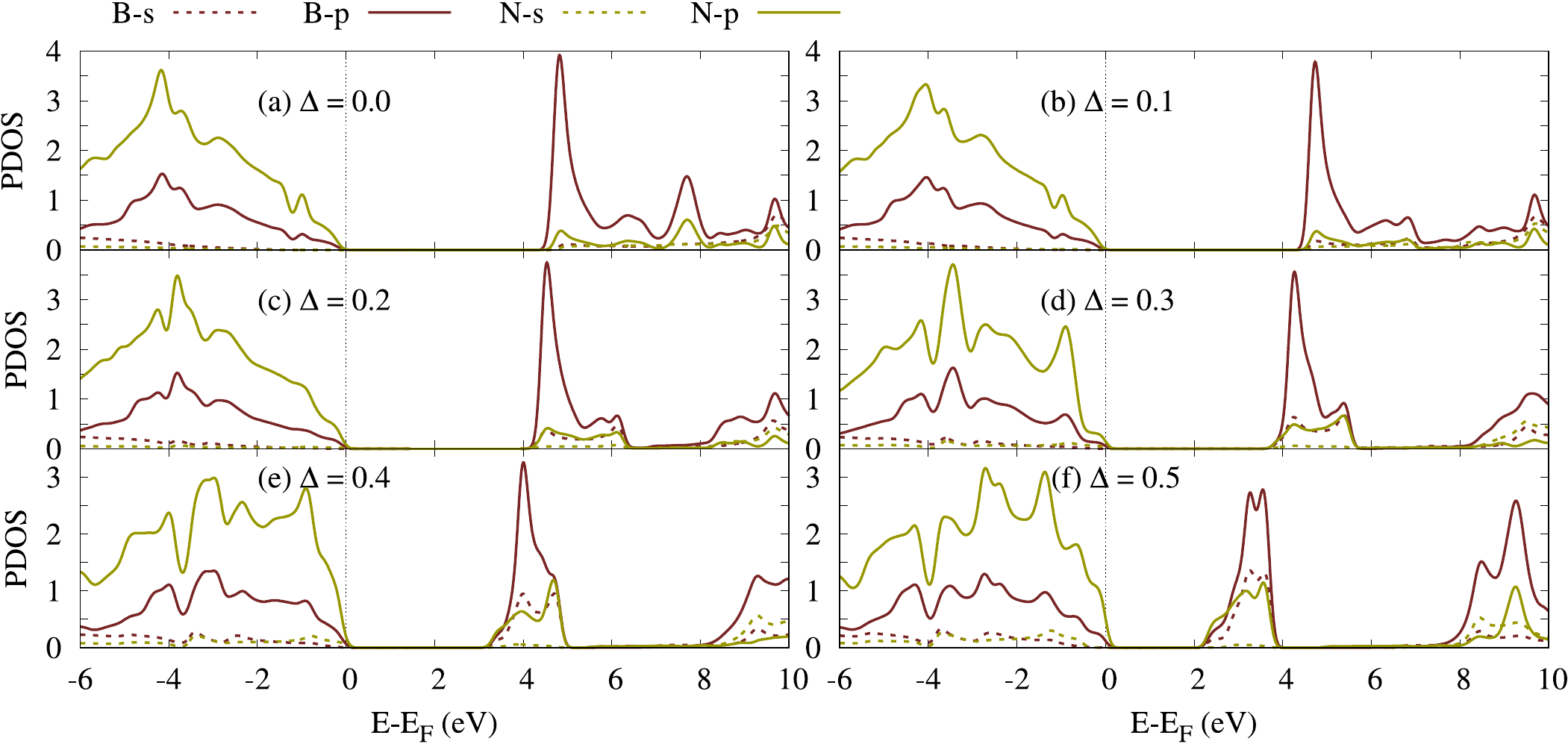}
	\caption{Partial density of states, PDOS of BN monolayers with planar buckling, $\Delta = 0.0$ (a), $0.1$ (b), $0.2$ (c), $0.3$ (d), $0.4$ (e), and $0.5$~$\angstrom$ (f). The energies are with respect to the Fermi level set to zero.}
	\label{fig02}
\end{figure*}
In addition, increasing planar buckling leads to elongation of the BN-bonds while the lattice constant is almost unchanged for all the considered values of buckling (see \tab{table_one}).
The elongation of the BN-bond influences the band structure, and the band gap E$_g$ is reduced with increasing planar buckling (this will be shown later).

The unit cell of BN consists of two different atoms, B and N atoms, and most BN layers are insulators or semiconductors with a wide band gap property. In the case of a flat or a planar BN monolayer, $\Delta = 0.0$, strong $\sigma$-bonds or covalent bond are generated via a sp$^2$ hybridization of overlapping orbitals \cite{Ci2010}.
The degree of the hybridization of the $s$- and the $p$-orbitals can be found using a simple equation,
$\cos(\theta) = s/(s-1) = (p-1)/p$, where $\theta$ is the angle between the equivalent orbitals, $s$ and $p$ \cite{kaufman1993inorganic}.
In \tab{table_two}, we demonstrate the contribution of the $s$- and the $p$-orbitals in the BN-bonds (in percentages) and hybridization order in the BN monolayer for all considered values of $\Delta$.
We can clearly see that the hybridization order is strongly influenced by the planar buckling, which causes the orbital redirection and bond reconstruction.
The sp$^2$ orbital hybridization is converted to an almost sp$^3$ hybridization due to the increased planar buckling. For instance, the hybridization becomes sp$^{2.93}$ at $\Delta = 0.5$~$\angstrom$. The planar buckling thus reduces the sp$^2$ overlapping, and the bond symmetry is broken simultaneously.
\begin{table}[h]
	\centering
	\begin{center}
		\caption{\label{table_two} Percentage of orbital contribution in the bond character and the hybridization bond order.}
		\begin{tabular}{|l|l|l|l|l|l|l|l|}\hline
			$\Delta$ ($\angstrom$)& $s$ ($\%$)  &  $p$ ($\%$) &  Hybrid      \\ \hline
			0.0	    & 33.33 & 66.66 &  sp$^2$      \\
			0.1	    & 32.97 & 67.03 &  sp$^{2.03}$ \\
			0.2	    & 32.02 & 67.98 &  sp$^{2.12}$ \\
			0.3	    & 30.45 & 69.55 &  sp$^{2.28}$ \\
			0.4	    & 28.21 & 71.79 &  sp$^{2.54}$ \\
			0.5	    & 25.4  & 74.6  &  sp$^{2.93}$ \\
			0.6	    & 21.9  & 78.1  &  sp$^{3.56}$ (not possible) \\
			0.7	    & 17.7  & 82.3  &  sp$^{4.64}$  (not possible) \\ \hline
	\end{tabular}	\end{center}
\end{table}
If the planar buckling is further increased to $0.6$ or $0.7$~$\angstrom$, the power of the $p$-orbital becomes greater than $3$, which is not possible because there is only one $s$ orbital and three $p$ orbitals. Therefore, we cannot have sp$^{>3}$ as there is no fourth $p$-orbital to hybridize.
So, we assume that $\Delta \leq 5$~$\angstrom$ in our calculations.

The effect of buckling on the band structure of BN monolayer is shown in \fig{fig01} for all considered values of planar buckling, and the band gap value is displayed in \tab{table_one}.
We have found that the band gap of a flat BN monolayer, $\Delta = 0.0$, is indirect with band gap value of $4.63$ eV along the $\Gamma$-K path, which agrees well with other DFT results \cite{Wickramaratne2018}.
The band gap is reduced when the planar buckling is increased, and it becomes a direct band gap from $\Delta = 0.4 \, \angstrom$ along the $\Gamma$ point.  It is clearly seen that increased planar buckling reduces the energy gap by shifting the conduction band minimum towards the valence band maxima or the Fermi level.
In addition, we notice that an increase in $\Delta$ is accompanied by a slight split of bands close to the valence band maxima and a strong split of bands close to the conduction band minima into discrete bands.
This leads to diminished degeneracy of energy levels resulting in a breaking the symmetry of the BN monolayer. Consequently, the $\Delta$ causes to a break the the sp$^2$ bonds symmetry, resulting in a reduction in the band gap.

In order to understand more details about the band structure and the energy gap reduction, the partial density of states (PDOS) is plotted in \fig{fig02} for various value of $\Delta$.
It is obvious that the valence density of states is formed by the $p$-orbitals of the B and the N atoms, which is similar for various value of $\Delta$, while the conduction density of states is more sensitive to changes in $\Delta$.
With increasing $\Delta$, the conduction density of states of the $s$- and the $p$-orbital for both the B and the N atoms move towards the Fermi energy resulting in a reduction of the semiconducting energy band gap.

We want to be more precise in determining which $p$-orbitals form the density of states.
In the case of a flat BN monolayer, $\Delta = 0.0$, we can confirm that the valence and conduction partial density of states near to the Fermi energy are formed due to a mix of only the $p_z$ orbitals of both the B and the N atoms (not shown).
In the presence of a planar bucking, for example for $\Delta = 0.5$~$\angstrom$, the contribution of $p_{x,y}$ in the PDOS is increased in addition to the $p_z$ orbitals of both the B and the N atoms in the valence band region, while the conduction partial density of states is formed by the $p_z$ orbitals of the B and the N atoms and the $s$-orbital of the B atoms.
This confirms that the planar buckling influences the $\sigma$ and the $\pi$ bonds of the BN monolayer leading to enhanced $\sigma\text{-}\pi$ common bonds in the density of states as well as the band structure.

\begin{figure}[htb]
	\centering
	\includegraphics[width=0.22\textwidth]{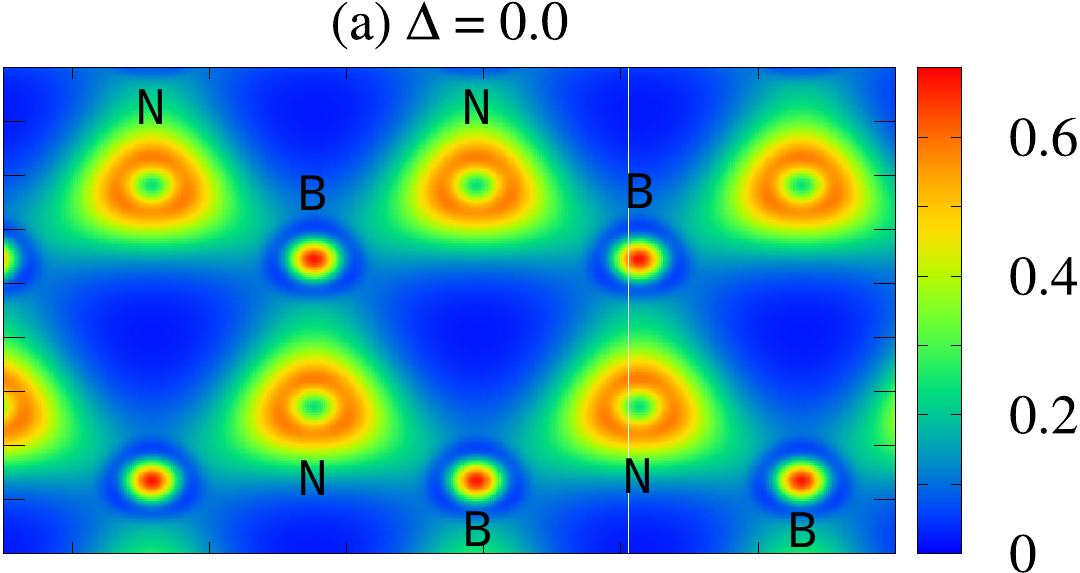}
	\includegraphics[width=0.22\textwidth]{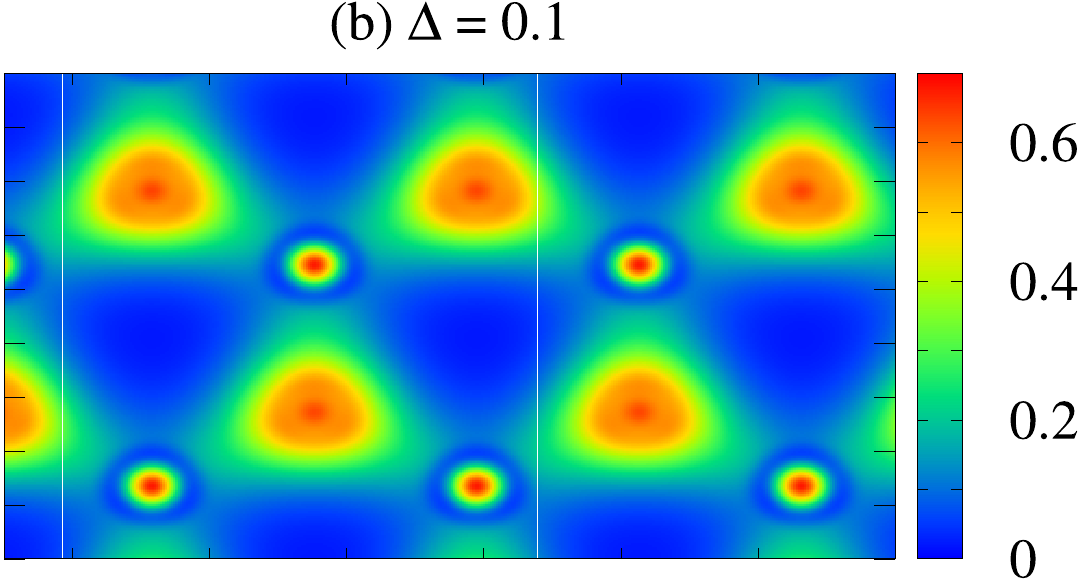}\\
	\includegraphics[width=0.22\textwidth]{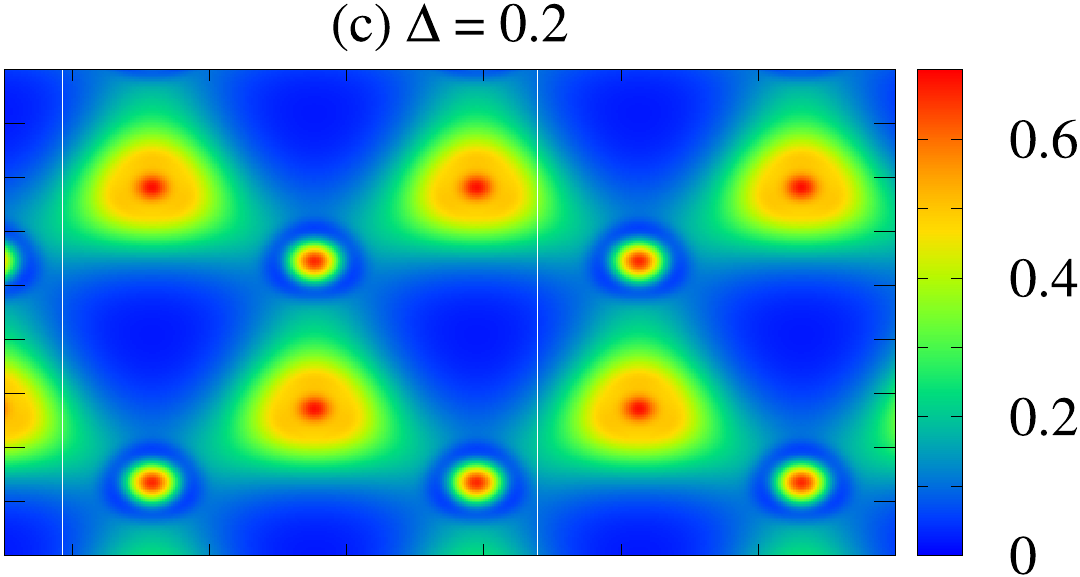}
	\includegraphics[width=0.22\textwidth]{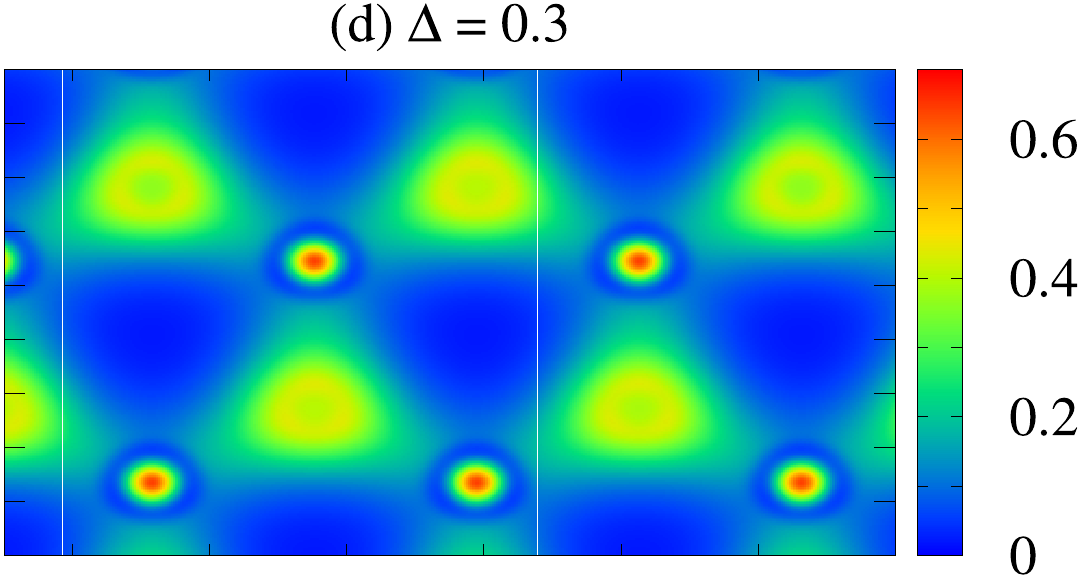}\\
	\includegraphics[width=0.22\textwidth]{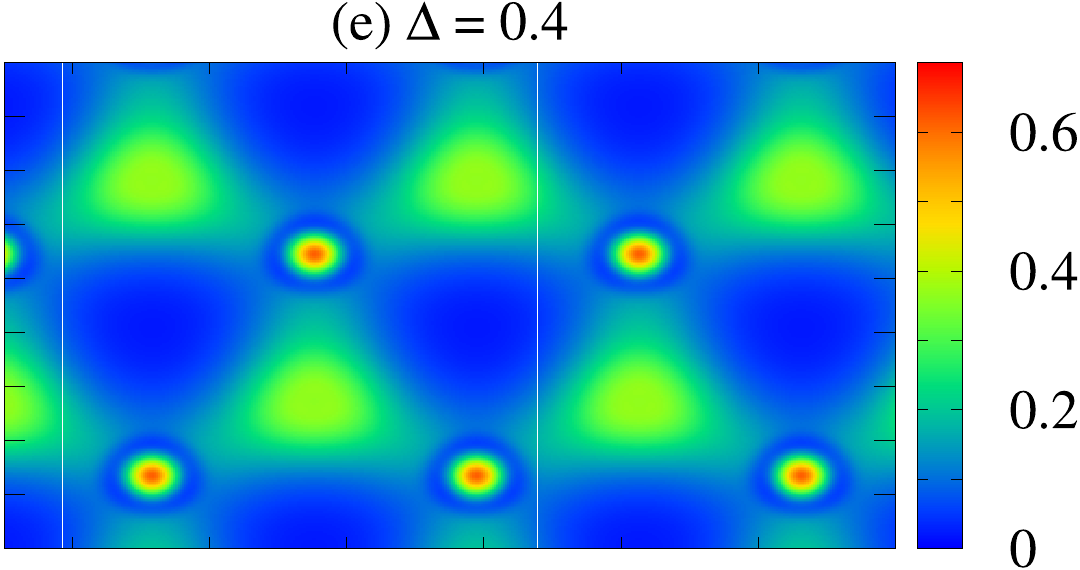}
	\includegraphics[width=0.22\textwidth]{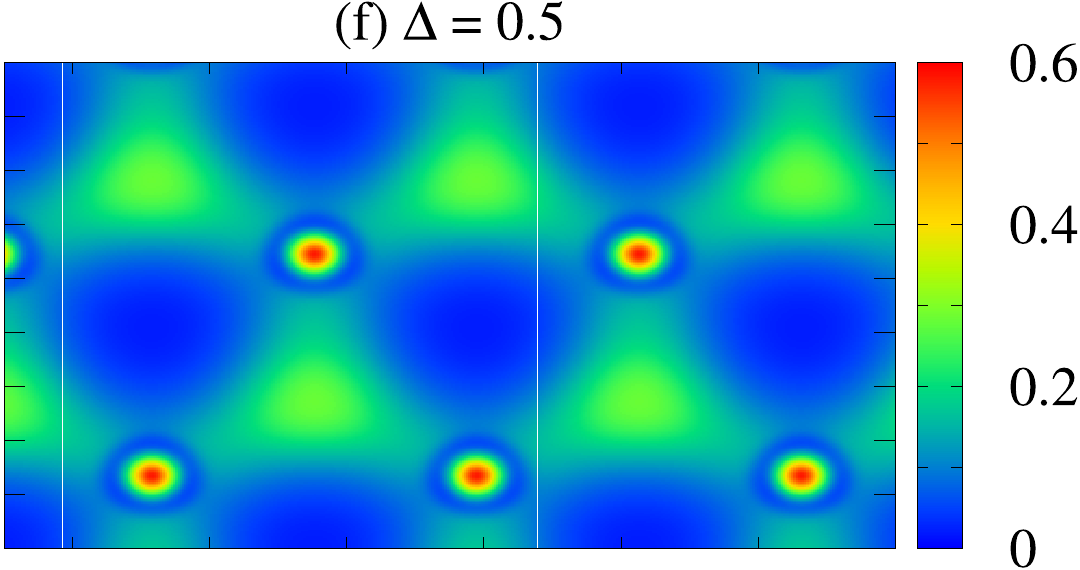}\\
	\caption{Electron density of BN monolayers with planar buckling, $\Delta = 0.0$ (a), $0.1$ (b), $0.2$ (c), $0.3$ (d), $0.4$ (e), and $0.5$~$\angstrom$ (f).}
	\label{fig03}
\end{figure}

The electron density distribution indicating valence electron density of a BN monolayer with different values of $\Delta$ is shown in \fig{fig03}. In the electron density of a planar BN monolayer, $\Delta = 0.0$, a high electron localization is found in the region between adjacent B and N atoms indicative of a strong $\sigma\text{-}\sigma$ covalent bonding. The electron density is also strongly localized around the N atoms, which is expected as the electronegativity of the N atom, $3.04$, is larger than that of the B atom, $2.04$ \cite{Ooi_2005}. This is a confirmation of a strong $\sigma$ bond between the B and the N atoms arising from the sp$^2$ hybridization.

The electron density distribution along the B and N bond depends on the value of $\Delta$, so that electron localization is broadened with increasing $\Delta$, and no significant overlap of the two neighboring atoms is seen for $\Delta = 0.5$~$\angstrom$. 
\lipsum[0]
\begin{figure*}[htb]
	\centering
	\includegraphics[width=0.9\textwidth]{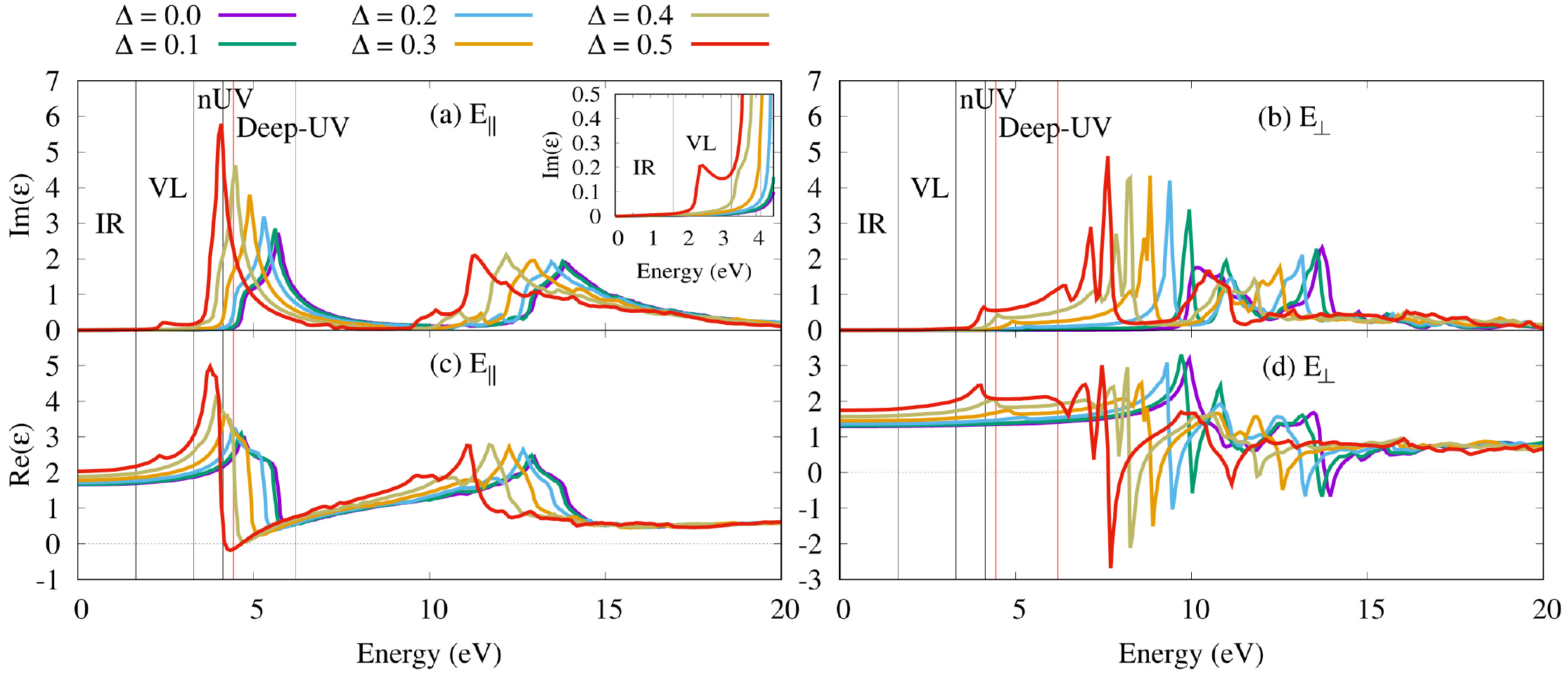}
	\caption{Imaginary, Im($\varepsilon$), (a,b) and real, Re($\varepsilon$), (c,d) parts of dielectric function for the BN monolayer with buckling parameter, $\Delta = 0.0$ (purple), $0.1$ (green), $0.2$ (light blue), $0.3$ (orange), $0.4$ (light brown), and $0.5$~$\angstrom$ (red) in the case of E$_{\parallel}$ (left panel), and E$_{\perp}$ (right panel). The vertical black and red lines indicate different regions of the electromagnetic spectrum.}
	\label{fig04}
\end{figure*}
This indicates that the $\sigma\text{-}\sigma$ covalent bond gets weaker with increasing $\Delta$ caused by forming of a sp$^3$ hybridization. The sp$^3$ hybridization forms a stronger $\sigma\text{-}\pi$ bond in the BN monolayer when $\Delta$ is increased.

\subsection{Optical properties}
Our next aim is to see the influences of the planar buckling on the optical properties
of a BN monolayer.
In the calculations of the optical properties including the imaginary part of the dielectric function,
the random phase approximation (RPA) has been used in which a very dense mesh grid, $100 \times 100 \times 1$, in the Brillouin zone is used to obtain accurate results \cite{PhysRev.115.786}.

The imaginary part of the dielectric function (a,b), Im($\varepsilon$), and the real part of the dielectric function (c,d), Re($\varepsilon$), are presented in \fig{fig04} for parallel (left panel) and perpendicular (right panel) polarization of the incoming electric field. Several regions in the Im($\varepsilon$), and Re($\varepsilon$) spectra are highlighted via vertical lines including the infrared, IR, ($0\text{-}1.65$~eV), the visible regime, VL, ($1.65\text{-}3.3$~eV), the near ultraviolet, nUV, ($3.3\text{-}4.13$~eV), and the Deep-UV regions ($4.428\text{-}6.199$~eV). The two red vertical lines show the Deep-UV region.

The Im($\varepsilon$) shows many significant variations with respect to a flat
BN monolayer, $\Delta=0.0$. In the case of a parallel electric field, E$_{\parallel}$, a sharp peak for a flat BN monolayer is seen centered at $5.69$~eV in the Deep-UV region. The location of this peak corresponds to the optical band gap of a flat BN monolayer, which agrees well with optical band gap measured experimentally \cite{Evans_2008} and predicted theoretically \cite{BEIRANVAND2015190} in previous studies. The peak becomes stronger and it shifts toward low energy ranges of the nUV region when $\Delta$ increases.
The intense peak remains in the Deep-UV region for several values of planar buckling up to $\Delta = 0.4$~$\angstrom$. In addition, an extra peak indicating the electronic band gap is seen when $\Delta = 0.5$~$\angstrom$ in the VL region (inset). These phenomena come from the shifting of the $s$- and the $p$-orbitals of the N atom and the $p$-orbital of B the atoms towards lower energy with creasing $\Delta$ in the conduction band region (shown \fig{fig02}). This shifting generates more states closer to the Fermi levels. As a result, the first intense peak indicating the transition from the valence to the conduction band is dislocated to lower energy ranges.
In the case of a perpendicular electric field, E$_{\perp}$, no significant effect in Im($\varepsilon$) is seem in the highlighted regions except that a small peak indicating the optical band gap moves from the Deep-UV to the nUV for increasing values of $\Delta$.

The same scenario is seen for Re$(\varepsilon)$ except that the first strong peak appears at $4.73$~eV in the Deep-UV region for a flat BN monlayer, $\Delta = 0.0$, and it moves to the nUV region for $\Delta = 0.5$~$\angstrom$ with a higher intensity when E$_{\parallel}$ is considered.
In addition, a small peak in the Re$(\varepsilon)$ spectra is formed from the Deep-UV to the the nUV regions in the case of E$_{\perp}$ in addition to the main intense peak. All of these phenomena in Re($\varepsilon$) arise from the variation of the $\sigma\text{-}\sigma$ bond or the in-plane bond nature in which the $\sigma$ bond gets weaker and moves towards a side-by-side overlapping, $\sigma\text{-}\pi$ bond, when $\Delta$ is increased.
The Re$(\varepsilon)$ spectra show an anisotropy behavior for different polarizations of the applied electric field.

The static dielectric constant (Re($\varepsilon(0)$)), the real part of the dielectric function at the zero value of energy, is another interesting parameter to study. It becomes a bit larger when the planar buckling is increased. The value of Re($\varepsilon(0)$) is $1.66$ (E$_{\parallel}$), and $1.30$ (E$_{\perp}$) for a flat BN monolayer, $\Delta = 0.0$, and these values are increased to $2.03$ (E$_{\parallel}$) and $1.74$ (E$_{\perp}$) for  $\Delta = 0.5$~$\angstrom$. It has been shown that the value of Re($\varepsilon(0)$) is inversely proportional to the band gap, Re$(\varepsilon(0)) \approx 1/E_{g}$ \cite{PhysRev.128.2093}, which can be used to verify our results of Re$(\varepsilon(0))$.
As the value of $\Delta$ is increased, the band gap is reduced resulting in an increase of Re$(\varepsilon(0))$. We note that the increased value of Re($\varepsilon(0)$) for E$_{\perp}$ is higher than that for E$_{\parallel}$, which again indicates the anisotropic behavior of Re($\varepsilon(0)$).
\begin{figure}[htb]
	\centering
	\includegraphics[width=0.45\textwidth]{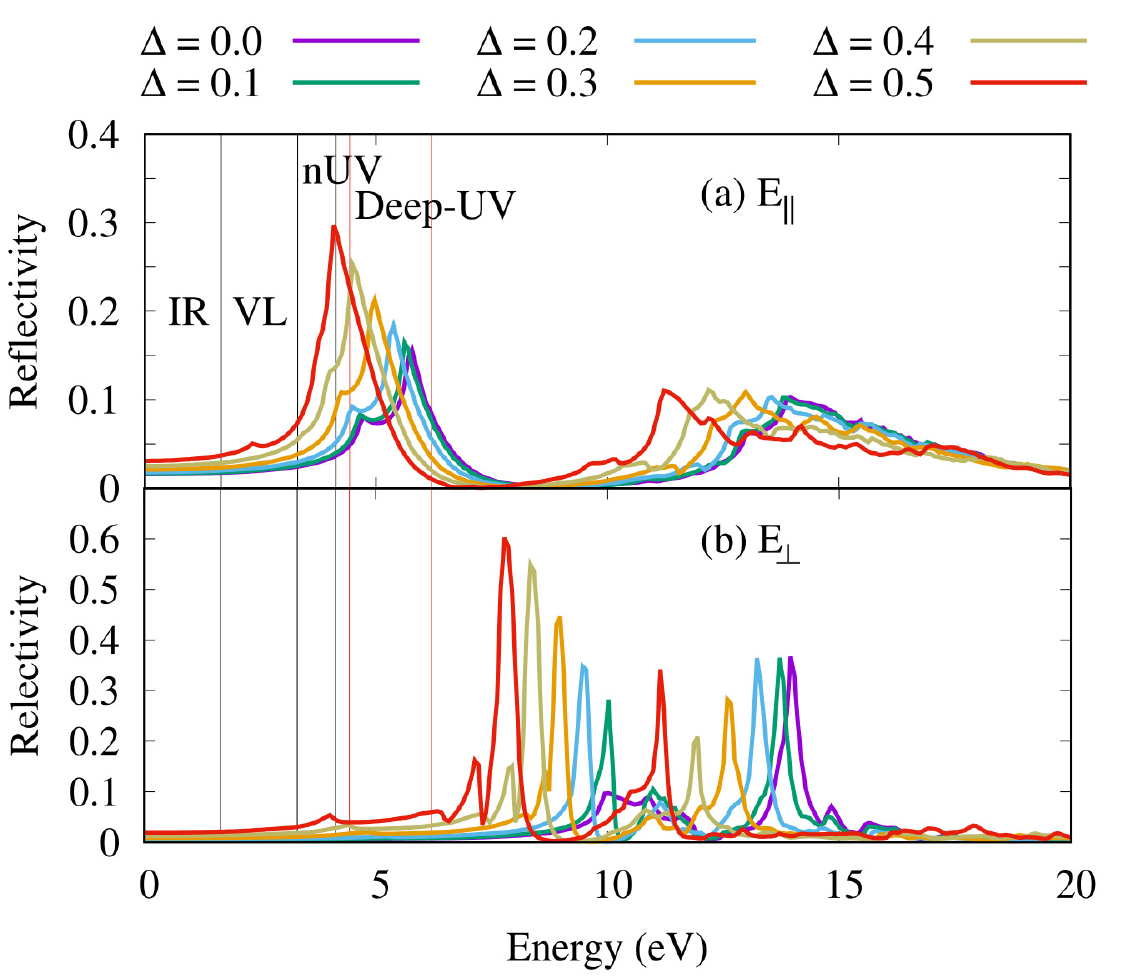}
	\caption{Reflectivity spectra of the BN monolayer with buckling parameter, $\Delta = 0.0$ (purple), $0.1$ (green), $0.2$ (light blue), $0.3$ (orange), $0.4$ (light brown), and $0.5$~$\angstrom$ (red) in the case of E$_{\parallel}$ (a), and E$_{\perp}$ (b). The vertical black and red lines show different regions of the electromagnetic spectrum.}
	\label{fig05}
\end{figure}

The materials transparency behavior is interesting to investigate as it is very useful for making coatings that provide new properties to materials.
It has been shown that the flat BN monolayer is optically transparent \cite{Li_2017}, and the transparency can be tuned by the planar buckling.
The effects of the planar buckling on reflectivity is presented in \fig{fig05} for E$_{\parallel}$ (a), and E$_{\perp}$ (b).
It is found that for E$_{\parallel}$ to the monolayer, the reflectivity
at smaller energy is more and at this energy range, the IR and the VL regions, the transmission is less (shown in \fig{fig04}(a)).
In our calculations, the reflection percentage of a flat BN monolayer in the VL range of the electromagnetic spectrum is $2.0\%$ and $0.5\%$ for E$_{\parallel}$ and E$_{\perp}$, respectively.
These reflectivity values are enhanced to  $5.0\%$ and $2.5\%$ for E$_{\parallel}$ and E$_{\perp}$, respectively, when the value of the planar buckling is increased to $0.5$~$\angstrom$. This confirms that the transparency in the VL range decreases as the planar buckling increases.
On the other hand, the reflectivity of a flat BN monolayer is $15.65\%$ in E$_{\parallel}$ forming a peak centered at the energy value of $5.77$~eV in the Deep-UV region. The reflectivity peak becomes sharper with an intensity $29.73\%$ and is dislocated to lower energy range about $4.08$~eV in the nUV region.
This indicates that the transparency from the Deep-UV to the nUV is decreases with increasing $\Delta$.
The modification in reflectivity by tuning the planar buckling is mostly caused by a rearranged electron density as is seen in \fig{fig03}.
The electron density rearrangement occurs by weakening the the $\sigma\text{-}\sigma$ bonds at high values of $\Delta$ leading to formation of stronger $\sigma\text{-}\pi$ bonds in a buckled BN monolayer.

\begin{figure}[htb]
	\centering
	\includegraphics[width=0.45\textwidth]{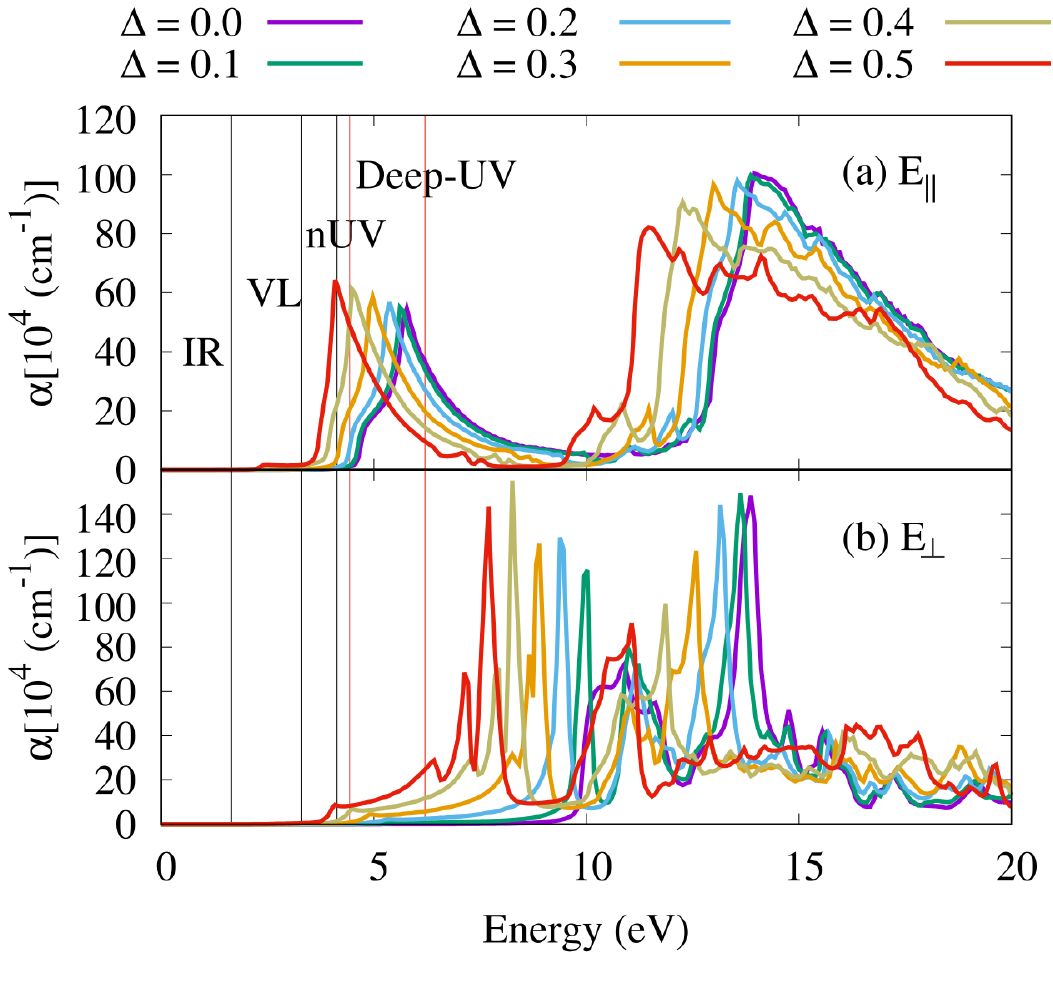}
	\caption{Absorption coefficient of a BN monolayer with a buckling parameter, $\Delta = 0.0$ (purple), $0.1$ (green), $0.2$ (light blue), $0.3$ (orange),  $0.4$ (light brown), and $0.5$~$\angstrom$ (red) in the case of E$_{\parallel}$ (a), and E$_{\perp}$ (b). The vertical black and red lines show different regions of the electromagnetic spectrum.}
	\label{fig06}
\end{figure}

The next target is to investigate the absorption coefficient, $\alpha$. The value of $\alpha$ is proportional to the sum over the interband transitions from the valence to the empty conduction energy states over the first Brillouin-zone k-points.
A BN monolayer with higher $\alpha$ more readily absorbs photons, which excite electrons into the conduction band. So, the $\alpha$ spectra will follow the Im($\varepsilon$) spectra shown in \fig{fig04}(a,b). The absorption spectra for both E$_{\parallel}$ (a), and E$_{\perp}$ (b) are presented in \fig{fig06}.

One can clearly see an enhancement of the absorption coefficient when the planar buckling is increased in the case of E$_{\parallel}$.
For instance, the absorption rate for a BN monolayer with $\Delta = 0.4$~$\angstrom$ is increased by $(12\text{-}15)\%$ comparing to a flat BN monolayer in the Deep-UV region.
If the planar buckling is further increased to $0.5$~$\angstrom$, the absorption peak is shifted to the nUV region with an absorption rate of $(15\text{-}20)\%$. A small absorbance peak is found in the VL region. Similar to Im($\varepsilon$) spectra, no significant change in the $\alpha$ spectra for the highlighted regions is induced for E$_{\perp}$.
The properties of the $\alpha$ spectra for E$_{\parallel}$ have a potential to boost the performance of various opto-electronic BN devices in the UV region such as deep-UV communications,
UV photodetectors, and UV spectroscopy, similar to a suggestion of a recent study of potential UV applications of graphene monolayers \cite{Cheng2021}.

\begin{figure}[htb]
	\centering
	\includegraphics[width=0.45\textwidth]{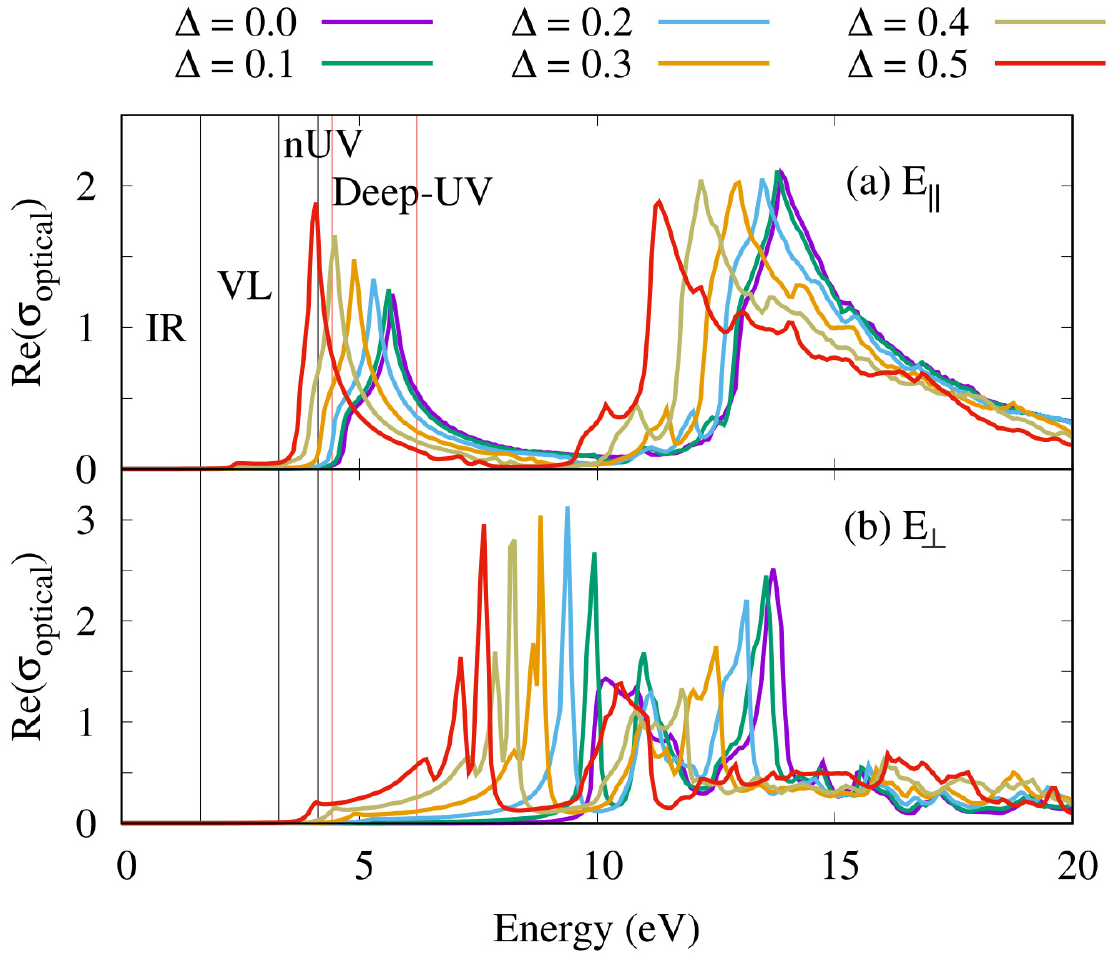}
	\caption{Optical conductivity (real part) of a BN monolayer with buckling parameters, $\Delta = 0.0$ (purple), $0.1$ (green), $0.2$ (light blue), $0.3$ (orange), $0.4$ (yellow), $0.4$ (light brown), and $0.5$~$\angstrom$ (red) in the case of E$_{\parallel}$ (a), and E$_{\perp}$ (b). The vertical black and red lines show different regions of the electromagnetic spectrum.}
	\label{fig07}
\end{figure}

The final result of our study is the optical conductivity, $\sigma_{\rm optical}$, of a BN monolayer with different values of $\Delta$ for E$_{\parallel}$ (a), and E$_{\perp}$ (b) shown in \fig{fig07}.
In fact, the real part of the optical conductivity, Re($\sigma_{\rm optical}$) is presented implying that the conductivity comes from an optical
excitation. As a gradient in the concentration of carriers in different regions of a semiconductor BN monolayer exists, the value of the optical conductivity becomes non-zero above the location of the band gap.
Above the band gap, a flat BN monolayer begins to have strong optical dispersion as a consequence of the interband transitions from the $s$- and the $p$-orbitals of the N atoms to the $s$- and the $p$-orbitals of the B atoms. As a result, the first strong peak is found at $5.69$~eV for E$_{\parallel}$ in the Deep-UV region, and the peak is shifted down to the nUV with a higher intensity when the planar buckling is increased just as the imaginary part of the dielectric function. The optical band gap can thus be easily recognized from Re($\sigma_{\rm optical}$), which is dislocated to the lower energy ranges.

\section{Conclusions}\label{Sec:Conclusion}
The electronic properties such as the band structure, the density of states, the electron density, and the optical behaviors such as the dielectric function, the reflectivity, the absorption, and the optical conductivity are displayed in this work for flat and buckled hexagonal BN monolayers.
The calculations were done based with a GGA-PBE method using density functional theory with a fully relaxed BN monolayer.
The generated planar buckling in the BN monolayer can tune it's wide band gap to a smaller band gap leading to modifications of the valence and tje conduction density of states in which the sp$^2$ hybridization of a flat BN monolayer is changed to a sp$^3$ hybridization in the presence of a planar buckling.
On the other hand, the planar buckling gives rise weaker $\sigma\text{-}\sigma$ and $\pi\text{-}\pi$ covalent bonds and makes stronger $\sigma\text{-}\pi$ bonds.
All these phenomena modify the optical properties in which the most intensive peaks in the imaginary part of the dielectric function, the absorption spectra, and the optical conductivity appear in the Deep-UV region, and remain in the same region with a higher intensity when the planar buckling is increased. Consequently, the optical properties can be enhanced by tuning the planar buckling in the Deep-UV region.
These robust features may be useful for optoelectronic and biosensing applications working in the UV region. In addition, a sign of the optical enhancement at a bit higher planar buckling is seen in the visible light region, which can be used in optoelectronic devices.

\section{Acknowledgment}
This work was financially supported by the University of Sulaimani and
the Research center of Komar University of Science and Technology.
The computations were performed on resources provided by the Division of Computational
Nanoscience at the University of Sulaimani.



\end{document}